\documentclass{eas}
\usepackage{graphicx}
\begin{document}

\title{Do open clusters have distinguishable chemical signatures?}
\runningtitle{Do open clusters have distinguishable chemical signatures?}
\author{S. Blanco-Cuaresma}\address{CNRS / Univ. Bordeaux, LAB, UMR 5804, F-33270, Floirac, France.}
\author{C. Soubiran$^{1}$}
\author{U. Heiter}\address{Department of Physics and Astronomy,  Uppsala University, Box 516, 75120 Uppsala, Sweden.}
\begin{abstract}
    Past studies have already shown that stars in open clusters are chemically homogeneous (e.g. De Silva et al. 2006, 2007 and 2009). These results support the idea that stars born from the same giant molecular cloud should have the same chemical composition. In this context, the chemical tagging technique was proposed by Freeman \& Bland-Hawthorn 2002. The principle is to recover disrupted stellar clusters by looking only to the stellar chemical composition. In order to evaluate the feasibility of this approach, it is necessary to test if we can distinguish between stars born from different molecular clouds. For this purpose, we studied the chemical composition of stars in 32 old and intermediate-age open clusters, and we applied machine learning algorithms to recover the original cluster by only considering the chemical signatures.
\end{abstract}
\maketitle
\section{Introduction}

Hundreds to thousands of stars can be formed in aggregates from one single molecular clouds
(Shu et al. \cite{1987ARA&A..25...23S}, Meyer et al. \cite{2000prpl.conf..121M}, Lada et al. \cite{2003ARA&A..41...57L}). 
If we assume that the progenitor cloud was chemically well-mixed (Feng et al. \cite{2014Natur.513..523F}),
then we expect to observe homogeneous chemical composition in the stars formed from this cloud
as already confirmed by De Silva et al. (\cite{2006AJ....131..455D}, \cite{2007AJ....133.1161D} and \cite{2009PASA...26...11D}).

Since each molecular cloud has its own history of pollution by ejecta from core-collapse supernovae (i.e. Type II, Ib and Ic Supernovae where most of the $\alpha$-element are produced), Type Ia supernovae (SNe Ia where most iron peak elements are created) and asymptotic giant branch stars (AGB where a s-process takes place), we expect different open clusters to have different chemical patterns. With this information, we could use the method of chemical tagging to track individual stars back to their common formation sites as proposed by Freeman \& Bland-Hawthorn (\cite{2002ARA&A..40..487F}).

The majority of the stars formed in clusters in our Galaxy have already been dispersed into the field, but a few of them managed to stay gravitationally bound. Old and intermediate-age open clusters (age $>\sim$100 Myr) represent a fantastic laboratory to investigate if we can distinguish between stars born from different molecular clouds. 

In this study, we collected high-resolution spectra of open clusters' stars observed by different instruments, we implemented a completely automatic process to derive atmospheric parameters and chemical abundances, and we used machine learning algorithms to try to recover the original clusters from the homogeneously derived chemical abundances.

\section{Observations}

We compiled more than 400 high-resolution spectra  of which 20\% come from the NARVAL instrument 
(Auri{\`e}re \cite{2003EAS.....9..105A}), 40\% from HARPS 
(Mayor et al. \cite{2003Msngr.114...20M}) and 40\% from UVES 
(Dekker et al. \cite{2000SPIE.4008..534D}). 
We mainly looked for clusters discussed on Paunzen et al. (\cite{2010A&A...517A..32P}) and Heiter et al. (\cite{2014A&A...561A..93H}), although we did not strictly limit the selection to these. To match the setup in the Gaia ESO Survey (Gilmore et al. \cite{2012Msngr.147...25G}), we chose to limit the spectral analysis to the range between 480 and 680~nm and we considered only spectra with a resolution of at least 47,000.

\section{Abundance determinations}

An automatic computational process was developed to derive atmospheric parameters and chemical abundances. The process is based on the integrated spectroscopic framework named iSpec (Blanco-Cuaresma et al. \cite{2014A&A...569A.111B}) and it was calibrated using the library of Gaia FGK benchmark stars (Blanco-Cuaresma et al. \cite{2014A&A...566A..98B}). The atomic data was kindly provided by the GES line-list sub-working group prior to publication (Heiter et al., in prep.). The line-list covers our wavelength range of interest and it also provides a selection of high quality lines (based on the reliability of the oscillator strength and the blend level), which are ideal for the determination of chemical abundances. As model atmosphere we used MARCS\footnote{http://marcs.astro.uu.se/} (Gustafsson et al. \cite{2008A&A...486..951G}) with the solar abundances from Grevesse et al. (\cite{2007SSRv..130..105G}).

\begin{figure*}
    \begin{centering}
        \includegraphics[width=6cm, trim = 8mm 8mm 5mm 5mm, clip]{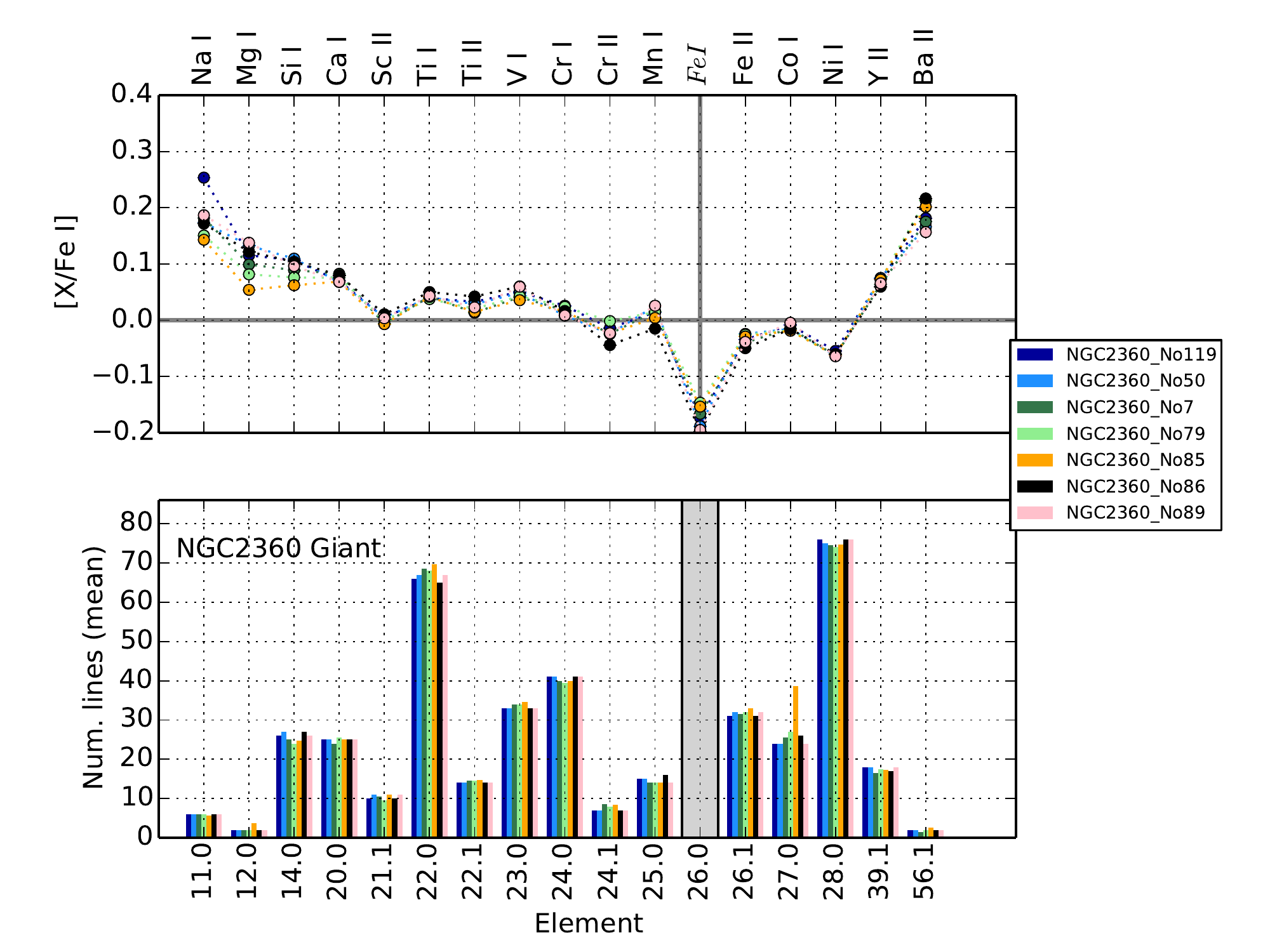}
        \includegraphics[width=6cm, trim = 8mm 8mm 5mm 5mm, clip]{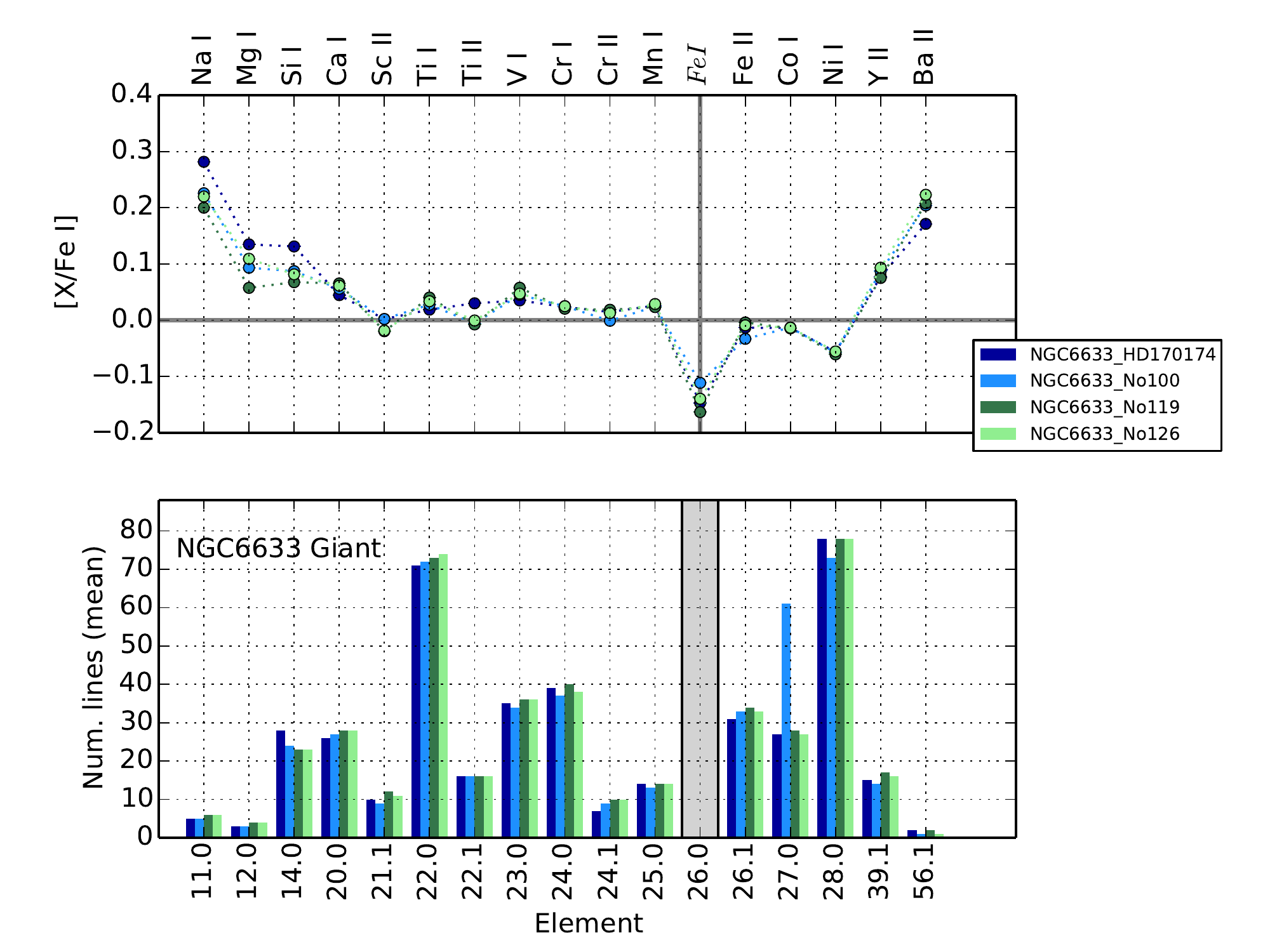}
        \par
    \end{centering}
    \caption{Abundances (top) and mean number of lines used (bottom) in function of species (element code at top, atomic number and ionization state in Kurucz format at bottom where '0' is neutral and '1' is ionized) for giant stars in NGC2360 and NGC6633. All the abundance ratios are referenced to iron except iron itself, which is relative to hydrogen. Each color represents a star with an identification name shown in the legend.}
    \label{fig:abundances}
\end{figure*}

After filtering non cluster members (validated by radial velocity), chemically peculiar and non-FGK stars, we were left with a dataset of more than 200 stars covering 32 open clusters with abundances for 17 species corresponding to 14 different elements (Fig. \ref{fig:abundances}).

\section{Recovering the original clusters}

To evaluate if we can distinguish stars born from different molecular clouds, we designed a blind experiment where we try to recover the original clusters using only the stellar chemical abundances. The stars were separated per evolutionary stage (i.e. dwarfs and giants) to reduce the impact of NLTE effects and the physical atomic diffusion process.

\begin{figure*}
    \begin{centering}
        \includegraphics[width=9cm, trim = 1mm 1mm 1mm 1mm, clip]{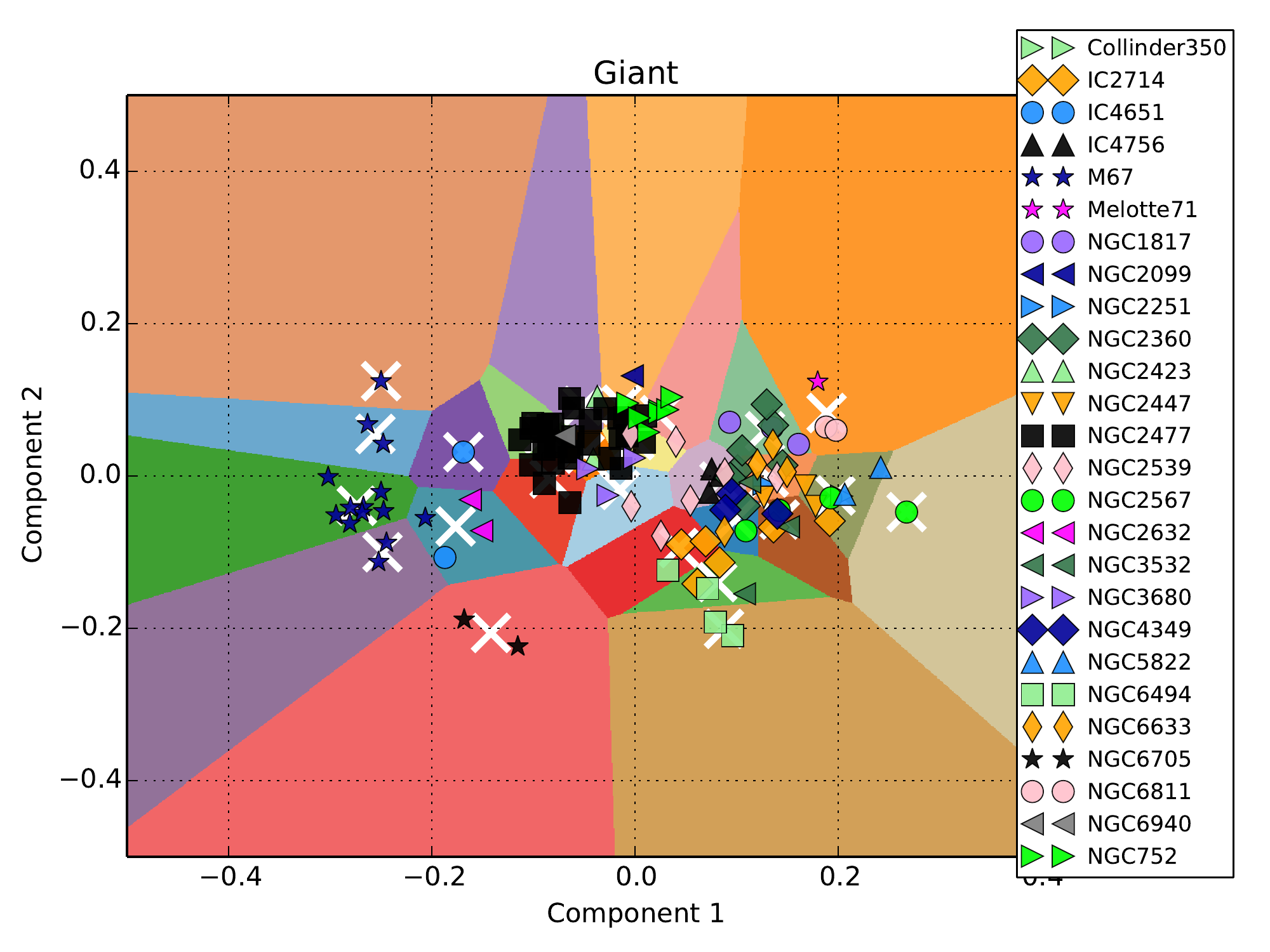}
        \par
    \end{centering}
    \caption{Giants represented using the first two components of PCA. Background colors correspond to the clusters found by the K-Means algorithm. Centroids are marked with white crosses.}
    \label{fig:pca}
\end{figure*}

We compressed the 17 abundances into two dimensions using the Principal Components Analysis (PCA), and we executed a K-Means clustering algorithm in order to form groups of stars chemically similar (Fig. \ref{fig:pca}). Even if the conditions of the experiment were extremely favorable (i.e. separation per evolutionary stage, filtered chemically peculiar stars), we were not able to recover the original clusters because of their high degree of overlapping in the chemical space. 

\section{Conclusions}

Given the set of elements used in this study, we found that some open clusters have very similar chemical signature (such as the clusters shown in Fig. \ref{fig:abundances}). Our blind experiment was not able to successful recover the original clusters using only the chemical abundances. Nevertheless, there are elements that seem to have a better discriminatory power such as the heavy n-capture element Ba. For future analysis, it would be interesting to include other elements such as La, Nd and Eu that are formed through similar processes that produce Ba (slow and rapid n-capture processes in low mass AGB, Brusso et al. \cite{2001ApJ...557..802B}; and core-collapse supernovae, Kratz et al. \cite{2007ApJ...662...39K}).


\end{document}